# Theory of Sound Propagation in Superfluid $He^3 - He^4$ Solutions Filled Porous Media

Sh.E.Kekutia, N.D.Chkhaidze.

Institute of Cybernetics, 5 Sandro Euli , Tbilisi, 0186, Georgia

Abstract: A theory of the propagation of acoustic waves in a porous medium filled with superfluid $He^3 - He^4$ solution is developed. The elastic coefficients in the system of equations are expressed in terms of physically measurable quantities. The equations obtained describe all volume modes that can propagate in a porous medium saturated with superfluid $He^3 - He^4$ solution. Finally, derived equations are applied to the most important particular case when the normal fluid component is locked inside a highly porous media (aerogel) by viscous forces and the velocities of two longitudinal sound modes are calculated.

## 1. INTRODUCTION

Sound propagation in a porous medium filled with a liquid has been investigated theoretically and experimentally for a long time. Especially, the investigation of porous media filled with superfluid helium is a rapidly developing research [1,2,3,4,5]. The biggest success in this field is stipulated by development of new techniques for producing impure superfluids, which exhibit unique properties. The introduction of the impurity particles into liquid helium produces impurity-helium (Im-He) clusters, which make it possible to create macroscopic Ym-He samples consisting of impurity atoms isolated in liquid or solid helium. At first these systems were obtained by injecting atoms and molecules such as nitrogen, deuterium, neon and krypton [6,7,8]. Superfluid helium confined to aerogel [9, 10], superfluid in Vycor glasses [11] and watergel-a frozen water "lattice" in HeII [12] can also be related to this new class of systems. Usually the macroscopic samples of the Im-He solid phase are built from aggregations of small Im-He clusters. Furthermore, these aggregates form extremely porous solids into which liquid helium can easily penetrate. These porous solids consist of a loosely or tightly connected continuous network of impurities or clusters of impurities each of which is surrounded by one or two layers of solidified helium. Therefore there is a unique opportunity to investigate the properties of superfluid helium in porous structures.

Below we discuss and consider the sound properties of above mentioned systems and show that sound phenomena in porous media filled with impure superfluids are modified from these in pure superfluid saturated porous media.

Namely, sound propagation in impure superfluids (superfluid $He^3 - He^4$ solution can be considered as an example of an impure superfluid) has a number of peculiarities connected with the oscillations of pressure and temperature in the acoustic wave. Whereas in pure helium II only the pressure oscillates in the first sound wave, and only the temperature oscillates in the second sound wave (neglecting the coefficient of thermal expansion, which is enormously small for helium), in a solution there are pressure, temperature, and concentration oscillations in both waves. In the first sound wave the oscillation of the temperature is proportional to the coefficient $\beta = (c/\rho) \partial \rho / \partial c$ and in the second sound wave the pressure oscillation is proportional to same coefficient (c- $He^3$ concentration, $\rho$ - density of the solution), and at low $He^3$ concentration the quantities proportional to $\beta$ can not be neglected ($\beta$ = -0,3-0,4 for highly concentrated solutions). Unlike pure $He^4$, the first sound wave in solutions contains a relative oscillation of the normal and superfluid liquids, the magnitude of which is proportional to $\beta$. In pure $He^4$, there are no oscillations of the total flux $\vec{J} = \rho^n \vec{V}^n + \rho^S \vec{V}^S$ in the second sound wave, whereas in the solution the deviation from the equilibrium value of $\vec{J}$ is also proportional to $\beta$ [13]. We note that here we imply dilute mixtures of different impurities (atoms, ions, molecules, etc) in superfluids. As an example, the dilute $He^3 - He^4$ mixtures could be considered as weak solutions of $He^3$ in $He^4$ because of the 6,4% maximum $He^3$ concentration. Sound phenomena in $He^3 - He^4$ mixtures have been investigated very intensively both theoretically and experimentally [14,15]. The theory of weak solutions was first developed by Khalatnikov [16]. We note that aerogel in superfluid has been studied very intensively during last decade [1, 2, 5, 9, 10, 17, 18, 19]. Aerogels are a unique class of ultra size, low density, open-cell foams. Aerogels have continuous porosity and a microstructure composed of interconnected colloidal-like particles or polymeric chains with characteristic diameters of 100 angstroms. The microstructure of aerogels is responsible for their unusual acoustic, mechanical, optical, and thermal properties. These microstructures impart high surface areas to the aerogels, for example, from about 350 m$^2$/g to about 2000 m$^2$/g. There ultra fine cell/pore size minimizes light scattering in the visible spectrum, and thus, aerogels can be prepared as transparent, porous solids. Further, the high porosity of aerogels makes them excellent insulators with their thermal conductivity being about 100 times lower than that of the prior known dense matrix foams. Still further, the aerogel skeleton provides for the low sound velocities observed in aerogels. Currently, aerogels of various compositions are known, and these aerogels were generally referred to as inorganic (such as silicon aerogels) and organic (such as carbon aerogels). Such inorganic aerogels, as silica, alumina, or zirconia aerogels, are traditionally made by means of the hydrolysis and condensation of metal alkoxides, such as tetrametoxy silane. Organic aerogels, such as carbon aerogels, are typically made from the sol-gel polymerization of resorcinol or melamine with formaldehyde under alkaline conditions. Therefore it is obvious that, when aerogel is saturated even with pure He II new phenomena are caused by the presence of aerogel: namely, the coupling between the two sound modes is provided by $\sigma \rho^a \rho^s$ ($\rho^a$ -is the aerogel density, $\sigma$ helium entropy) [17], which enhances the coupling between the two sound modes. Such replacement of coupling between sounds accompanies by change of sound conversion character in impure superfluids and for superfluids in aerogel. For example, the propagation of a heat pulse with the velocity of the first mode in HeII in aerogel has been observed [10].

Such unusual sound phenomena as slow pressure oscillations and fast temperature oscillations can also be observed in impure superfluids. So, as we see superfluidity of helium in restricted geometries has been the object of much theoretical and experimental interest in recent years. It is not a theory, that can completely describe the wave phenomena in solid porous media saturated with superfluid $He^3 - He^4$ solution. Thus the objective of the article represents the derivation of hydrodynamic equations for consolidated porous media filled with superfluid $He^3 - He^4$ solution and the determination of all input elastic coefficients of the theory by physically measured quantities. A porous material filled with superfluid solution simultaneously possesses the properties of elastic solid and superfluid solution and therefore it is expected observation novel sound phenomena.

We start by calculating the generalized coefficients of the theory without any additional adjustable parameters and by producing the general equations, which can predict all bulk modes that propagate in $He^3 - He^4$ solutions saturated porous media. From derivation equations, we consider the case of unrestricted geometry, case of the fourth sound and the most important particular case, when normal fluid component is locked inside a porous media by viscous forces. We calculate the velocities of propagating longitudinal waves in highly porous media (aerogel) filled with superfluid $He^3 - He^4$ solution. Finally we discuss and summarize results.

## EXPRESSION OF GENERALIZED COEFFICIENTS BY PHYSICALLY MEASURED QUANTITIES.

The elastic properties of a system containing a superfluid helium completely filling the pores were considered in [1,2], where methods for measurement of generalized elastic coefficients are described with jacketed and unjacketed compressibility tests in the case of a homogeneous and isotropic porous matrix. We consider the case, when impurities participate only in normal fluid flow [13]. In our case according to [1,20] the stress-strain relations are

$$\sigma_x = 2N e_x + A e + Q^S \varepsilon^S + Q^n \varepsilon^n$$
$$\sigma_y = 2N e_y + A e + Q^S \varepsilon^S + Q^n \varepsilon^n$$
$$\sigma_z = 2N e_z + A e + Q^S \varepsilon^S + Q^n \varepsilon^n$$
$$\tau_x = N\gamma_x, \tau_y = N\gamma_y, \tau_z = N\gamma_z. \qquad (1)$$
$$s' = Q^S e + R^S \varepsilon^S + R^{Sn} \varepsilon^n$$
$$s'' = Q^n e + R^n \varepsilon^n + R^{Sn} \varepsilon^S$$

where $\sigma_x, \sigma_y, \sigma_z$ and $\tau_x, \tau_y, \tau_z$ are normal and tangential forces acting on an element of the solid surface with the following orientation, $s'$ and $s''$ - forces acting on the solution part, which correspond to superfluid and normal components of superfluid solution.

Since the present system is a porous structure, we assume that the unit volume is much larger than the pore size. Therefore we determine the displacement vector $\vec{u}$ as the displacement of the solid averaged over a volume element. The average displacement vector $\vec{U}$ of the liquid part of the cube, which determines the fluid flow, can be determined in the same manner.

The average displacement vector of the solid has the components $u_x, u_y, u_z$ and that of the mixture $U_x^S, U_y^S, U_z^S, U_x^n, U_y^n, U_z^n$. The solid strain components are then given by

$$e_x = \frac{\partial u_x}{\partial x},\ e_y = \frac{\partial u_y}{\partial y},\ e_z = \frac{\partial u_z}{\partial y},\ \gamma_x = \frac{\partial u_y}{\partial z} + \frac{\partial u_z}{\partial y},\ \gamma_y = \frac{\partial u_x}{\partial z} + \frac{\partial u_z}{\partial x},\ \gamma_z = \frac{\partial u_x}{\partial y} + \frac{\partial u_y}{\partial x} \quad (2)$$

Due to two possible types of motion in He II the displacement of superfluid solution $\vec{U}$ breaks down into the sum of two parts

$$\vec{U} = \frac{\rho^S}{\rho}\vec{U}^S + \frac{\rho^n}{\rho}\vec{U}^n \quad (3)$$

corresponding to displacement of superfluid and normal components. Thus the strain in fluid is defined by the dilatation

$$\varepsilon = \frac{\rho^S}{\rho}\nabla\vec{U}^S + \frac{\rho^n}{\rho}\nabla\vec{U}^n \quad (4)$$

Since the superfluid and normal components cannot be physically separated in He II and it is meaningless to talk about whether individual atoms of the liquid belong to the superfluid or normal components, the following relation should take place:

$$Q^S\varepsilon^S + Q^n\varepsilon^n = Q\varepsilon \quad (5)$$

The coefficient A and N correspond to the well-known Lame coefficient in the theory of elasticity and are positive. The coefficient Q and R are the familiar Biot's coefficients [22]. $R^S(R^n)$ is a measure of the stress arising in the superfluid (normal) component when a unit volume of the system is compressed without compressing the normal (superfluid) component and the porous medium. The coefficient $R^{Sn}$ determines the stresses arising in the superfluid component when the normal component is compressed without compression of the superfluid component and the porous medium, and vice versa [1, 20].

Let us discuss hypothetical experiments which make it possible to relate generalized elastic coefficients of the theory in terms of the measured coefficients: the bulk modulus of fluid $K_f$, the bulk modulus of solid $K_{sol}$, the bulk modulus of the skeletal frame $K_b$ and $N$. Such an approach was proposed in [1, 21] for superfluid and normal liquid.

In the unjacketed compressibility experiment, a sample of the porous solid is immersed in a superfluid $He^3 - He^4$ solution to which a pressure $p'$ was applied. Under the action of pressure the solution penetrates the pores completely and the dilations of the porous solid e and solutions $\varepsilon$ are measured. Therefore $K_{sol}$ and $K_f$ can be determined as

$$\frac{1}{K_{sol}} = -\frac{e}{p'}; \quad \frac{1}{K_f} = -\frac{\varepsilon}{p'} \quad (6)$$

We also note, that from expression of the solution chemical potential we have the form of the force acting on the superfluid component and normal component portions:

$$s' = -\Phi \frac{\rho^S}{\rho}(1+\beta)p'; \quad s'' = -\Phi \frac{\rho^n}{\rho}\left(1 - \frac{\rho^S}{\rho^n}\beta\right)p' \quad (7)$$

Where $\Phi$ is porosity.

After the consideration of the conditions $\varepsilon^S = \varepsilon^n = \varepsilon$ we have

$$\left(\frac{2}{3}N + A\right)\frac{1}{K_{sol}} + \left(Q^S + Q^n\right)\frac{1}{K_f} = (1-\Phi)$$

$$Q^S \frac{1}{K_{sol}} + \left(R^S + R^{Sn}\right)\frac{1}{K_f} = \Phi \frac{\rho^S}{\rho}(1+\beta) \quad (8)$$

$$Q^n \frac{1}{K_{sol}} + \left(R^n + R^{Sn}\right)\frac{1}{K_f} = \Phi \frac{\rho^n}{\rho}\left(1 - \frac{\rho^S}{\rho^n}\beta\right)$$

The following test corresponds to the jacketed compressibility test, when a specimen of the material is enclosed in a thin impermeable jacket and is subjected to pressure $p'$ of the solution. The superfluid solution inside the jacket should pass through a tube communicated with the external reservoir to insure constant internal solution pressure. The dilatation of the specimen is measured and coefficient of jacketed compressibility $K_b$ is determined by

$$\frac{1}{K_b} = -\frac{e}{p'} \tag{9}$$

In this experiment

$$\sigma_x = \sigma_y = \sigma_z = -p'; \; \varepsilon^S = \varepsilon^n = \varepsilon; \; s' = s'' = 0 \tag{10}$$

Therefore we have three relations

$$\left(\frac{2}{3}N + A\right)e + \left(Q^S + Q^n\right)\varepsilon = -p'$$
$$Q^S e + \left(R^S + R^{Sn}\right)\varepsilon = 0 \tag{11}$$
$$Q^n e + \left(R^n + R^{Sn}\right)\varepsilon = 0$$

From (9) and (12) it follows

$$Q = \frac{\Phi K_{sol}\left(1 - \Phi - \frac{K_b}{K_{sol}}\right)}{1 - \Phi + \Phi\frac{K_{sol}}{K_f} - \frac{K_b}{K_{sol}}} \tag{12}$$

$$\frac{2}{3}N + A = K_{sol}\frac{(1-\Phi)\left(1 - \Phi - \frac{K_b}{K_{sol}}\right) + \Phi\frac{K_b}{K_f}}{1 - \Phi + \Phi\frac{K_{sol}}{K_f} - \frac{K_b}{K_{sol}}} \tag{13}$$

$$R^n + R^{Sn} = K_f \frac{\rho^n}{\rho}\left(\Phi - \frac{Q}{K_{sol}}\right)\left(1 - \frac{\rho^S}{\rho^n}\beta\right) \tag{14}$$

$$R^S + R^{Sn} = K_f \frac{\rho^S}{\rho}\left(\Phi - \frac{Q}{K_{sol}}\right)(1 + \beta) \tag{15}$$

Let us consider the situation when the jacket is communicated with reservoir by the superleak. Therefore only superfluid component pours into reservoir and we can write the following relations:

$$\left(\frac{2}{3}N+A\right)e+Q^S\varepsilon^S+Q^n\varepsilon^n=-(1-\Phi)p'$$
$$Q^S e+R^S\varepsilon^S+R^{Sn}\varepsilon^n=0 \quad (16)$$
$$Q^n e+R^n\varepsilon^n+R^{Sn}\varepsilon^S=-\Phi p'$$

In this compressibility test the relation between $\varepsilon^S$ and $\varepsilon^n$ can be obtained using the laws of conservation of mass and entropy for superfluid $He^3-He^4$ solution. Then we have

$$(\varepsilon^S-\varepsilon^n)\frac{\tilde{\sigma}^2}{\partial\sigma/\partial T}\frac{\rho^S}{\rho}=\frac{1+\beta}{\rho}p' \quad (17)$$

Where $\tilde{\sigma}^2=\bar{\sigma}^2+c^2\frac{\partial}{\partial c}\left(\frac{Z}{\rho}\right)\frac{\partial\sigma}{\partial T}$; $\bar{\sigma}=\sigma-c\frac{\partial\sigma}{\partial c}$.

The quantity $Z=\rho(\mu_3-\mu_4)$ is defined in terms of the chemical potentials $\mu_3,\mu_4$ for $He^3$ and $He^4$ in the solution.

These equations (16-17) together with (12-15) give:

$$R^{Sn}=\frac{\rho^S\rho^n}{\rho^2}(1+\beta)\left(1-\frac{\rho^S}{\rho^n}\beta\right)R-\frac{(\rho^S)^2\tilde{\sigma}^2 T\Phi}{\rho\,C_{He}} \quad (18)$$

$$R^n=\frac{(\rho^n)^2}{\rho^2}\left(1-\frac{\rho^S}{\rho^n}\beta\right)^2 R+\frac{(\rho^S)^2\tilde{\sigma}^2 T\Phi}{\rho\,C_{He}} \quad (19)$$

$$R^S=\frac{(\rho^S)^2}{\rho^2}(1+\beta)^2 R+\frac{(\rho^S)^2\tilde{\sigma}^2 T\Phi}{\rho\,C_{He}} \quad (20)$$

Where Biot-Willis coefficient R is equal to [21]

$$R=\frac{\Phi^2 K_{sol}}{1-\Phi+\Phi\frac{K_{sol}}{K_f}-\frac{K_b}{K_{sol}}} \quad (21)$$

and $C_{He}$ is the specific heat of the solution. We note once again that the coefficients $K_f$, $K_{sol}$, $K_b$ and $N$ are experimentally measurable quantities.

We shall use Lagrange's formalism to find the equation of motion of the system porous solid- superfluid $He^3-He^4$ solution. To construct the desired equations the kinetic energy of the system must be determined after the generalized coordinates are chosen. We assume that the physical point is a region of size much greater than the pore size but much smaller than the characteristic lengths in the problem ( for example, the wave-length when considering wave processes). For the generalized coordinates of the system we choose the nine components of the displacement vectors of the solution and solid, averaged over the volume of the physical point: $u_x, u_y, u_z, U_x^S, U_y^S, U_z^S, U_x^n, U_y^n, U_z^n$. Then equations for propagating waves are received by analogy with articles [1, 22]. So, we have the equations in vector form:

$$N\nabla^2 \vec{u} + (A+N)\,grad\, e + Q^S grad\, \varepsilon^S + Q^n grad\, \varepsilon^n = \frac{\partial^2}{\partial t^2}\left(\rho_{11}\vec{u} + \rho_{12}^S \vec{U}^S + \rho_{12}^n \vec{U}^n\right) +$$
$$+ bF(w)\frac{\partial}{\partial t}\left(\vec{u} - \vec{U}^n\right)$$
$$Q^S grad\, e + R^S grad\, \varepsilon^S + R^{Sn} grad\, \varepsilon^n = \frac{\partial^2}{\partial t^2}\left(\rho_{12}^S \vec{u} + \rho_{22}^S \vec{U}^S\right) \qquad (22)$$
$$Q^n grad\, e + R^n grad\, \varepsilon^n + R^{Sn} grad\, \varepsilon^S = \frac{\partial^2}{\partial t^2}\left(\rho_{12}^n \vec{u} + \rho_{22}^n \vec{U}^n\right) - bF(w)\frac{\partial}{\partial t}\left(\vec{u} - \vec{U}^n\right)$$

Here we took into account that the main mechanism of dissipation in the system porous solid- superfluid $He^3 - He^4$ solutions is deceleration of the normal component of the superfluid liquid by the walls of the pores. Since all deformations are assumed to be small, the macroscopic motions studied in the theory are small elastic oscillations or waves. Consequently, as in most physical systems, the friction forces are proportional to the velocities of the moving physical points and can be described using a dissipative function. The dissipative function is, by definition, a homogeneous quadratic function of generalized velocities. The complex function F (w) reflects the deviation from a Poiscuille flow taking account of the geometric features of the porous material [23]; the coefficient $b = \eta \Phi^2 / k_0$ is the ratio of total friction force to the average normal fluid velocity, where $\eta$ is the viscosity of the fluid and $k_0$ is the permeability. The possibilities of using the acoustics of a superfluid liquid to study various parameters of porous materials were analyzed theoretically in Ref. 23. Finally the function F(w) was expressed in terms of the key parameters: the sinuosity, the permeability, a dynamical parameter with the dimension of length, and the porosity. Some of them were obtained from the solution of the problem of the electric conductivity of a porous medium consisting of an insulating porous material filled with a conducting liquid. The response of the rigid porous medium was calculated. The results obtained make it possible to investigate the characteristic features of two-fluid hydrodynamics of He II in a rigid porous medium and to determine the parameters of the medium from experimental data for the velocities of first, second, and fourth sounds.

In equations (22) $\rho_{11}$ is total effective density of the solid moving in the $He^3 - He^4$ solution. Coefficients $\rho_{12}^S$ and $\rho_{12}^n$ are mass parameters of "coupling" between a solid and correspondingly, superfluid and normal components of solution or mass coefficient $\rho_{12}^{S(n)}$ describes the inertial (as opposed to viscous) drag that the fluid exerts on the solid as the latter is accelerated relative to the former and vice-versa [1, 22, 24].

## SOUND PROPAGATION IN UNRESTRICTED GEOMETRY AND AEROGEL.

Now it will be interesting to ignore dissipative process in equations (23) and consider the case of unrestricted geometry. Then from equations (23) we have

$$R^S \, grad \, \varepsilon^S + R^{Sn} \, grad \, \varepsilon^n = \rho^S \frac{\partial^2 \vec{U}^S}{\partial t^2}$$

$$R^n \, grad \, \varepsilon^n + R^{Sn} \, grad \, \varepsilon^S = \rho^n \frac{\partial^2 \vec{U}^n}{\partial t^2}$$
(23)

Here we take into account that in the limit interest to us purely geometrical quantity $\alpha_\infty$, which is independently of solid or fluid densities, and porosity $\Phi$ are equal to one. Because the induced mass tensor per unit volume
$\rho_{12}^{S(n)} = -(\alpha_\infty - 1) \, \Phi \rho^{S(n)}$
[24] and

$$R^S = \frac{(\rho^S)^2}{\rho^2}(1+\beta)^2 K_f + \frac{(\rho^S)^2 \tilde{\sigma}^2 T}{\rho \, C_{He}}$$

$$R^n = \frac{(\rho^n)^2}{\rho^2}\left(1 - \frac{\rho^S}{\rho^n}\beta\right)^2 K_f + \frac{(\rho^S)^2 \tilde{\sigma}^2 T}{\rho \, C_{He}} \quad (24)$$

$$R^{Sn} = \frac{\rho^S \rho^n}{\rho^2}(1+\beta)\left(1-\frac{\rho^S}{\rho^n}\beta\right)K_f - \frac{(\rho^S)^2}{\rho C_{He}}\tilde{\sigma}^2 T$$

So, for pure $He^3 - He^4$ solution solving the system (23) in the usual manner we obtain the dispersion equation for the bulk waves propagating in free $He^3 - He^4$ solution:

$$C^4 \rho^S \rho^n - C^2(\rho^S R^n + \rho^n R^S) + R^S R^n - (R^{Sn})^2 = 0 \qquad (25)$$

Equation (25) has two roots:

$$C_1^2 = \frac{K_f}{\rho}\left(1+\frac{\rho^S}{\rho^n}\beta^2\right); \quad C_2^2 = \frac{\rho^S}{\rho^n}\frac{\tilde{\sigma}^2 T}{C_{He}\left(1+\frac{\rho^S}{\rho^n}\beta^2\right)} \qquad (26)$$

which conform to the velocity of the first and the second sounds correspondingly [16, 25, 26]. From (23) equations it follows the well-known results for the fourth sound in free $He^3 - He^4$ solutions [13]. If we assume $\vec{U}^n = 0$ in (23), we derive [25, 26]

$$C_4^2 = \frac{\rho^S}{\rho}C_1^2\frac{(1+\beta)^2}{1+\frac{\rho^S}{\rho^n}\beta^2} + \frac{\rho^n}{\rho}C_2^2\left(1+\frac{\rho^S}{\rho^n}\beta^2\right) \qquad (27)$$

Propagation of the fourth sound in a $He^3 - He^4$ solution was studied in [25, 26] from the Khalatnikov hydrodynamic equations.

A great deal of effort has recently been dedicated to the investigation of superfluid solution in porous materials. We cite here recent articles describing the specific features of superfluid liquid in various porous structures [27]. The sound velocity in porous media can provide information about both the superfluidity and elastic properties of the solid matrix. Mckenna at al [9] developed a theory explaining the behavior of sound modes in aerogel filled with He II, taking into account coupling between the normal component and the aerogel and its elasticity. Here the normal component is locked in a very compliant solid matrix so that the liquid and the aerogel fibers move together under mechanical and thermal gradients. It takes place at low sound frequencies, when the viscous penetration depth is bigger than the pore size so the entire normal component is viscously locked to the solid matrix. We discuss and consider the same phenomena in superfuid $He^3 - He^4$ solution-aerogel. In this case from (22) for longitudinal waves we have the following dispersion equation:

$$\rho^S[\rho^a + \rho^n]C^4 - C^2[R^S(\rho^a + \rho) + \rho^S(A + 2N + 2Q + R) - 2\rho^S \times$$
$$\times (Q^S + R^S + R^{Sn})] + R^S(A + 2N + 2Q + R) - (Q^S + R^S + R^{Sn})^2 = 0 \qquad (28)$$

For an aerogel or, equivalently, for an open geometry $\Phi \approx 1$ and $K_b << K_{sol}$. Under these conditions, taking into account the fact that in a "framework" which is not filled with liquid the velocity of the longitudinal wave $C_a^2 = \dfrac{K_b + (4/3)N}{\rho^a}$, the dispersion equations (28) becomes

$$\left(1 + \dfrac{\rho^a}{\rho^n}\right)C^4 - C^2\left\{C_1^2 + \left(1 + \dfrac{\rho^S}{\rho^n}\beta^2\right)C_2^2 + \dfrac{\rho^a}{\rho^n}\left(C_a^2 + C_4^2\right)\right\} + C_1^2 C_2^2 + \dfrac{\rho^a}{\rho^n}C_a^2 C_4^2 = 0 \quad (29)$$

The first solution is intermediate between the first and fourth sound

$$C_{14}^2 = \dfrac{C_1^2 + \dfrac{\rho^a}{\rho^n}C_4^2}{1 + \dfrac{\rho^a}{\rho^n}} \qquad (30)$$

and it resembles the fast mode.

Another solution corresponds to the slow mode, which is an oscillation of a deformation of the aerogel combined with a simultaneous out-of-phase motion of the superfluid component:

$$C_{2a}^2 = \dfrac{C_2^2 + \dfrac{\rho^a \rho^S}{\rho^n \rho}\dfrac{(1+\beta)^2}{1 + \dfrac{\rho^S}{\rho^n}\beta^2}C_a^2}{1 + \dfrac{\rho^a \rho^S}{\rho^n \rho}\dfrac{(1+\beta)^2}{1 + \dfrac{\rho^S}{\rho^n}\beta^2}} \qquad (31)$$

Second mode has been observed in HeII in aerogel as heat pulse propagation with the velocity of first sound. It should be observed on impure superfluids-aerogel also From experiment date for silica aerogel $C_a^2 >> C_2^2$ [9], so from the above mentioned formula it follows that $C_{2a}^2 >> C_2^2$ Therefore, the velocity of slow wave is much bigger than the velocity of temperature sound in free solutions.

From (30) and (31) it follows that an aerogel filled with superfluid $He^3 - He^4$ solution simultaneously possesses the properties of elastic solid and superfluid liquid. Also, in this article we have considered the peculiarities of sound propagation for impure homogeneous superfluids, where these phenomena are caused both by impurities (including $He^3$ in He II) and by the presence of aerogel.

In summary, we have obtained the hydrodynamic equations for the system porous body- superfluid $He^3 - He^4$ solution. The equations obtained describe all volume modes that can propagate in a porous medium saturated with superfluid $He^3 - He^4$ solution. The elasticity coefficients appearing in the equations were expressed in terms of physically measurable quantities. Derived equations were applied to the most important particular case when the normal fluid component is locked inside a highly porous media (aerogel) by viscous forces and the velocities of two longitudinal sound modes were calculated.

## L I T E R A T U R E